\documentclass[epsf]{aa}
\input{psfig}

\def\e{{\rm e}}
\def\d{{\rm d}}
\def\ie{i.e. }
\def\eg{e.g. }
\def\etal{et al. }
\def\cc{\rm c.c.}

\def\sist{\sigma_{\rm var}}
\def\somme{\Upsilon}

\newlength{\largeur}
\newlength{\saut}
\def\marge#1{
\setlength{\largeur}{\columnwidth}
\addtolength{\largeur}{-#1}
\setlength{\saut}{0.5\largeur}\hspace*{\saut}}
\def\picture #1 by #2 (#3){
 \marge{#1} \vbox to #2{
  \hrule width #1 height 0pt depth 0pt
  \vfill
  \special{picture #3}}}

\begin{document}

\thesaurus{9 (06.15.1, 06.13.1, 03.13.1, 03.13.6)} 

\title{Are solar acoustic modes correlated ?} 

\author{T. Foglizzo$^{1}$, R.A. Garc\'\i a$^{1}$, 
P. Boumier$^{2}$, J. Charra$^{2}$, A.H. Gabriel$^{2}$, G. Grec$^{3}$, 
J.M. Robillot$^{4}$, T. Roca Cort\'es$^{5}$, S. Turck-Chi\`eze$^{1}$, 
R.K. Ulrich$^{6}$}

\offprints{foglizzo@cea.fr}

\institute {
    $^1$ Service d'Astrophysique, DAPNIA/DSM, CE-Saclay, 91191 
Gif-sur-Yvette, 
France\\
	$^2$ Institut d'Astrophysique Spatiale, Orsay, France\\
	$^3$ Observatoire de la C\^ote d'Azur, Lab. Cassini CNRS URA1362,
 06304 Nice, France\\
	$^4$ Observatoire de l'Universite Bordeaux 1, BP 89, 33270 Floirac, 
France\\
	$^5$ Instituto de Astrof\'\i sica de Canarias,
             E-38205 La Laguna, Tenerife, Spain \\
	$^6$ UCLA, Department of Physics and Astronomy, Los Angeles, USA}

\date{9th September 1997}

\maketitle

\markboth{T. Foglizzo, R.A. Garc\'\i a \etal: 
Are solar acoustic modes correlated ?}{}

\begin{abstract}
We have studied the statistical properties of the energy of individual
acoustic modes, 
extracted from 310 days of GOLF data near the solar minimum. The 
exponential  distribution of the energy of each mode is clearly seen.
The modes are found to be uncorrelated with a  $\pm 0.6\% $ accuracy, 
thus supporting the hypothesis of sto\-chas\-tic excitation by the 
solar 
convection.\\ 
Nevertheless, the same analysis performed on the same modes just 
before the solar maximum, using IPHIR data, rejects the hypothesis of 
no correlation at a 
$99.3\%$ confidence level.\\
A simple model suggests that $31.3\pm 9.4\% $ of the energy of each 
mode is 
coherent among the modes studied in IPHIR data, correponding to a 
mean correlation of $10.7\pm 5.9\%$.
\keywords{Sun: oscillations; magnetic fields - methods: analytical, 
statistical}

\end{abstract}

\section{Introduction}
The quality of GOLF data offers a unique opportunity to investigate 
the excitation mechanism of low degree p modes. In the region of 4-6 
minutes, several modes can be extracted and analysed one by one with 
a short enough time resolution ($\sim 1.4$ days) over one year, 
allowing an accurate statistical treatment of the data obtained.\\
Here we deal with the ``superficial'' energy per unit of mass 
$E(t)/M$, 
associated with the mode at the surface of the sun. We do not address 
the issue
of the  ``global'' energy of the mode, which is related to the 
superficial
energy  through the shape of the eigenfunction inside the sun.\\
Woodard (1984) first revealed the exponential nature of the 
distribution of 
spectral power in each frequency bin, using ACRIM data.\\
Goldreich \& Keeley (1977) first considered the excitation of p modes
by turbulent motions near the surface of the convection zone. 
Using the analogy of a damped oscillator excited stochastically,
Kumar, Franklin \& Goldreich (1988) described analytically how the 
theoretical distribution of energy, averaged over a given 
time-window, should depend on 
the damping time of the mode. Comparisons with real data were 
performed 
by Toutain \& Fr\" ohlich (1992) with  160 days of IPHIR data. 
They found in particular that the damping times deduced from the 
linewidths were compatible with those expected in the  model of 
stochastic 
excitation.\\
Chang (1996) pointed out that strong localized peaks in the energy 
variations
of a damped oscillator excited sto\-chas\-ti\-ca\-lly do not 
necessarily correspond 
to a strong excitation, but rather to an exceptional coherent 
addition of 
random phases.\\
This is true provided that the number of independent excitations
per damping time of the mode is large enough, like in the case of 
excitation by
solar granules. In this context, two different modes excited by 
exactly the 
same sources would have uncorrelated energies.\\
Different modes, however, would be correlated, if the ti\-mes\-ca\-le 
between two excitations from a common source were longer than their 
damping time.\\
The correlation between the modes energies is therefore directly 
related to the characteristics of their source of excitation.\\
Using 160 days of IPHIR data at the end of 1988 (just before 
the  solar maximum, $\sim 1990$), Baudin \etal (1996) concluded that 
the
p modes were likely to be correlated. An anticorrelation between the 
mean 
solar magnetic field and the p-mode power 
was found by Gavryusev \& Gavryuseva (1997) in IPHIR data, while 
no clear correlation has been detected yet in GOLF data (Baudin \etal 
1997).\\
After checking in Sect.~\ref{sectexpo} the exponential nature of the 
distribution of energy of p modes in GOLF data, we address 
the issue of their relative independence in Sect.~\ref{sectindep},
using statistical tests based on Montecarlo simulations.\\
These same tests are used to re-analyse IPHIR data in 
Sect.~\ref{sectiphir}.

\section{Time evolution of the energy of a single 
mode\label{sectexpo}} 
\subsection{Method of extraction of the energy\label{extractm}}
The energy integrated over a time interval, \ie the power of the 
mode, was computed by Chaplin \etal (1995) using a Fourier transform 
over 
short  subseries. More sophisticated methods based on the wavelet 
analysis
were developped by Baudin, Gabriel \& Gibert (1994) in order to 
analyse the variations of power both with time and frequency.\\
Frequency resolution is not required for our study. Since the 
distribution of energy is likely to be mathematically simpler than 
the 
distribution of power (Kumar, Franklin \& Goldreich 1988), we have 
prefered to extract the energy directly.\\
Let $v(t)$ be the oscillatory velocity (\eg integrated over the 
surface of the sun), filtered in the Fourier domain through two 
windows of 
width $\Delta\omega$ centred on the eigenfrequencies $\pm\omega_{0}$.
Its Fourier transform ${\hat v}(\omega)$ is therefore equal to zero
out of these windows. The time evolution of the energy of this 
isolated mode 
can be obtained by a bivariate spectral analysis, as in 
Toutain \& Fr\" ohlich (1992). Here we favour
a simpler method based on the inverse Fourier transform $f_{v}(t)$ 
of the line, translated around $\omega=0$. It is shown in Appendix A
that the energy of this mode can be written as follows:
\begin{eqnarray}
f_{v}(t)&\equiv&\int_{-{\Delta \omega\over 2}}^{+{\Delta \omega\over 
2}}
{\hat v}(\omega_{0}+\omega)\e^{i\omega t}\d \omega,\label{defg}\\
{E\over M}(t)&=&2|f_{v}(t)|^{2}\left\{1+
{\cal O}\left({\Delta \omega\over\omega_{0}}\right)\right\}.
\label{energ}
\end{eqnarray}	
This approach is equivalent to the one used by Chang \& Gough (1995), 
by means of the Hilbert transform $H(v)$ of the velocity, since
$4|f_{v}(t)|^{2}\equiv v^{2}(t)+H(v)^{2}(t)$.\\
If the distribution of velocities is gaussian, then the real and 
imaginary 
parts of the function $f_{v}(t)$ are two independent 
gaussian  distributions with identical amplitudes and variances. Thus 
Eq.
(\ref{energ}) directly implies that the distribution of the energy is
exponential, as expected.\\ 
Eq.~(\ref{defg}) shows clearly that the time resolution $\delta t$ 
of  the
energy, reconstructed by Eq.~(\ref{energ}), is related to the size 
$\Delta\nu\equiv \Delta\omega/2\pi$ of the filtering window: 
\begin{equation}
\delta t = {1\over \Delta\nu}. \label{resol}
\end{equation}
Denoting by $T$ the total length of the observation, the 
frequency resolution of the Fourier transform is $1/T$, and the 
filtering 
window  $\Delta\nu$ contains $p\equiv T\Delta\nu $ points. The inverse
FFT  algorithm is used to compute Eq.~(\ref{defg}) and define the 
energy 
at $p$ successive instants. Eq.~(\ref{resol}) then guarantees that 
the 
resulting energy is not oversampled.

\subsection{Application to the GOLF data\label{sectevol}} 
\begin{figure} 
\psfig{file=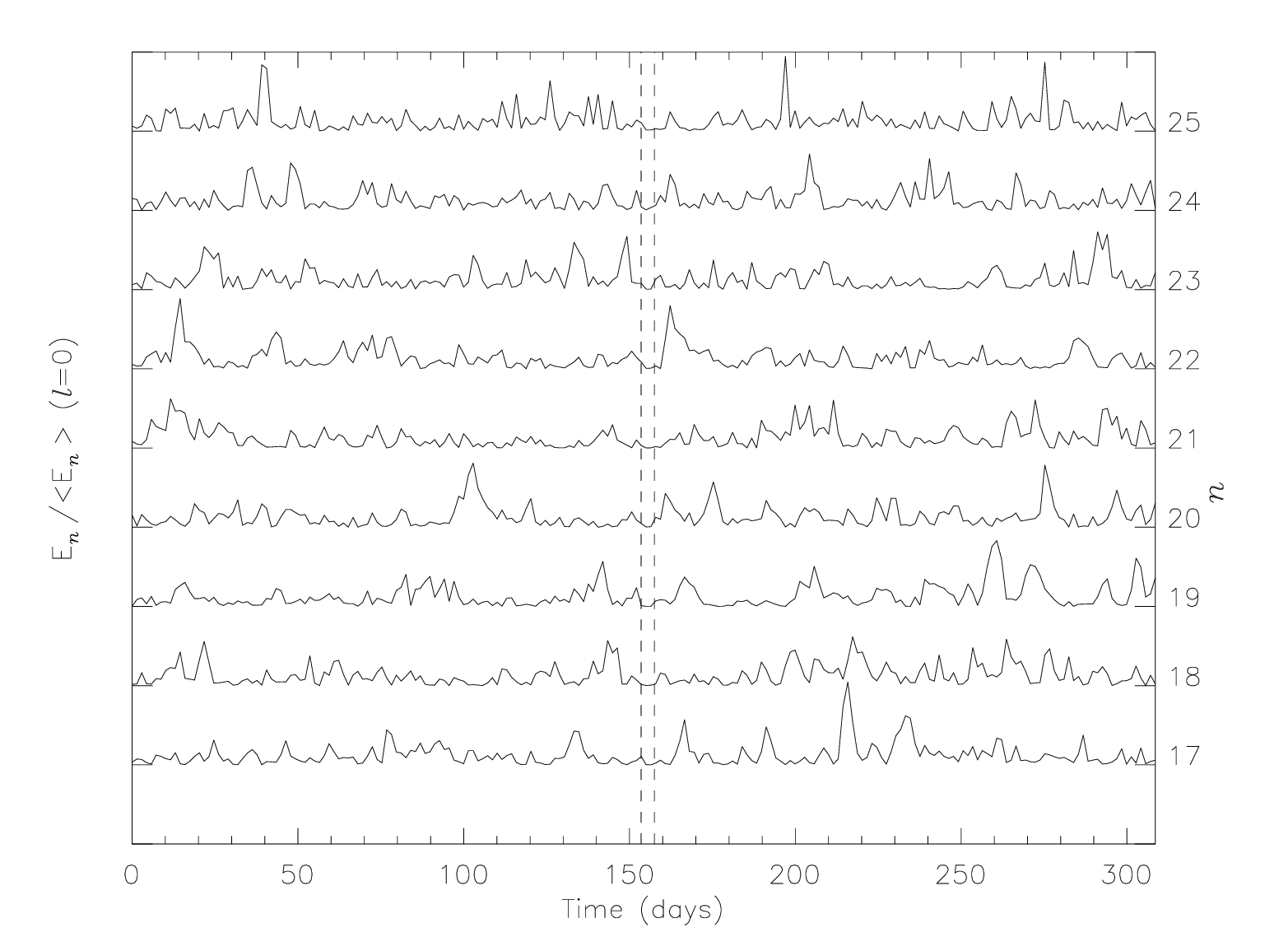,width=\columnwidth}
\psfig{file=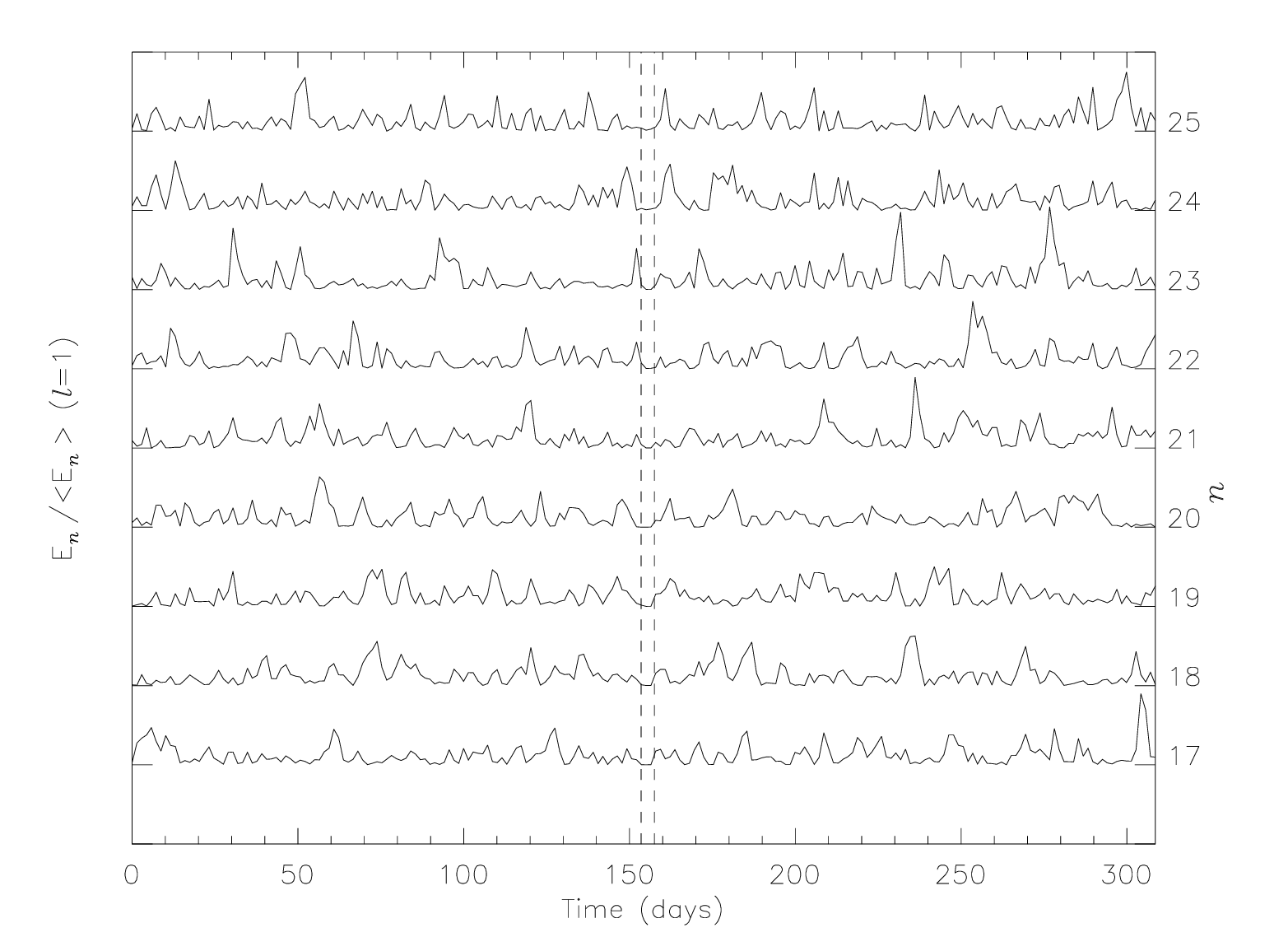,width=\columnwidth}
\caption[]{Time evolution of the energy of the modes $17\le n \le 25$,
$l=0$ (above) and $l=1$ (below). 
The energy of each mode is normalized to its mean value.}        
\label{etl01}    
\end{figure}
We have considered the set of p modes corresponding to $17\le n \le 
25$, 
$l=0$ and $1$, between 11th April 1996 and 14th February 1997 
(a publication concerning the calibration procedure is in 
preparation). 
The Fourier transform of 
the resulting velocity over these 310 days allows a filtering window 
size of 
$\Delta\nu=8\mu Hz$ ($\delta t\sim 1.45$ days) for this set of modes. 
The window is symmetric with respect to the centroid of the 
line, $\omega_{0}$, which is determined according to Lazrek \etal 
(1997). The two $m$-components of the mode $l=1$, however, are not 
separated. 
In contrast with IPHIR, the width of the window is 
determined by the proximity of another mode ($l=2,3$), rather than by 
the level 
of noise which is here very low.\\
Fig.~\ref{etl01} shows the time evolution of the energy of 
the 18 selected modes $l=0$ and $l=1$, normalized to their mean 
energy. The GOLF instrument was stopped 
for one day on 8th September 1996. Four days of signal were removed 
from our statistical study (around the 156th day on Fig.~\ref{etl01}) 
in order to account  for the stabilisation of the instrument. The 
resulting 
sample is made up of 210  points.
\subsection{Statistical tests\label{statest}}
\begin{figure} 
\psfig{file=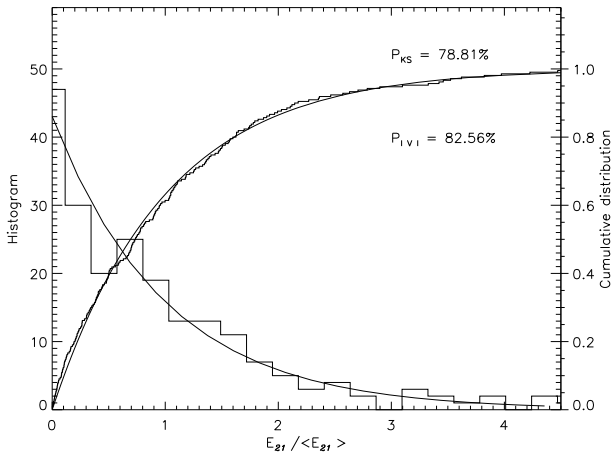,width=\columnwidth}
\caption[]{Histogram (20 bins) of the energy of the mode $l=0$, 
$n=21$, and 
its cumulative distribution, compared to a theoretical exponential
distribution. The variance test $P_{|V|}$ compares the observed 
variance to the 
theoretical one, while the Kolmogorov-Smirnov test $P_{KS}$ depends 
on the 
maximum distance between the theoretical and the observed cumulative 
distributions.}
\label{de22}    
\end{figure}

Following the picture of a thermodynamic equilibrium between the 
random motions of the convective cells and the oscillating cavity 
(Goldreich \& Keeley 1977), we wish to 
compare the observed sample of energies $\xi_{i},\;1\le i\le p$, with 
an 
exponential distribution. Any exponential distribution is defined by a
single parameter, its mean value $m$. Fig.~\ref{de22} shows a typical
histogram and cumulative distribution for the modes extracted from 
the GOLF
data (the cumulative distribution is defined as the primitive of the 
density of probability, it increases monotonously from 0 to 1). They 
are
compared to an exponential distribution whose mean value $m_{p}$ is
estimated from the sample of $p$ points. Using the Maximum Likelihood
approach, the best unbiased estimator of  $m$ for an exponential 
distribution is
the following: 
\begin{equation}
m_{p}\equiv {1\over p}\sum_{i}\xi_{i}.
\end{equation}
\par (i) The variance test\\
A simple test consists in checking that the first moments of the 
distribution (mean value and variance) are compatible with those of a 
theoretical exponential distribution.\\
The variance $\sigma^{2}$ of an exponential distribution coincides 
with the 
square of  its mean value. We check this property by computing, for 
each mode
of the  GOLF data, the ratio $V_{p}$ of the 
estimated variance (denoted by $\sigma_{p}^{2}$) to the estimated 
mean 
value squared $m_{p}^{2}$: 
\begin{equation}
V_{p}\equiv {
{1\over p-1}\sum_{j} \left(\xi_{j}-{1\over p}\sum_{i} 
\xi_{i}\right)^{2}
\over
\left({1\over p}\sum_{i}\xi_{i}\right)^{2}
}.
\end{equation}
Each value is interpreted owing to the cumulative distribution
$P^{p}_{V}$ of $V_{p}$, obtained if $V_{p}$ were built from a true 
exponential 
distribution. $P^{p}_{V}$ is computed numerically using a Montecarlo 
method, with $10^{5}$ exponential samples of $p$ points. 
For each of the modes selected, $P^{p}_{V}$ is the fraction 
of these $10^{5}$ trials leading to a value of $V_{p}$ 
{\it larger} than the one observed. Since we are interested 
only in knowing whether the observed $V_{p}$ is typical of an 
exponential distribution or not, we shall give equal importance to 
the 
lowest and highest values of the variance by measuring
the quantity $P_{|V|}\equiv 2\;{\rm min}(P_{V},100-P_{V})$.
\par (ii) The Kolmogorov-Smirnov test\\
While the variance test depends only  on one particular moment of the 
observed distribution $\xi_{i}$, a more global comparison is achieved 
with 
the Kol\-mo\-go\-rov-Smir\-nov (KS) test on the cumulative 
distribution $S_{p}(\xi)$.
This test measures the maximum distance $d(S_{p},P_{m})$ between 
$S_{p}(\xi)$ and a theoretical exponential cumulative distribution
$P_{m}(x)\equiv 1 - \e^{-x/m}$. 
If the mode energies were exponentially distributed, the statistics 
of $d(S_{p},P_{m})$ would be described by a cumulative distribution
denoted by $P^{p}_{KS}$. Since the mean value $m$ of the reference 
ditribution 
is estimated from the data, we cannot use the standard formulae 
(Numerical Recipies 1992, Chapt. 14.3) to fit $P^{p}_{KS}$.\\
Instead of doing this, we have used a Montecarlo method of $10^{5}$ 
samples in 
order to define the cumulative distribution $P^{p}_{KS}$ of the 
distance 
$d(S_{p},P_{m_{p}})$. $P^{p}_{KS}$ therefore indicates the fraction 
of these $10^{5}$ trials leading to a distance $d(S_{p},P_{m_{p}})$
{\it larger} than the value observed.\\
For each of the modes selected, a value of $P^{p}_{KS}$ close
to $0\% $ would indicate that the observed distribution is too far 
from the 
theoretical one. A value of $P^{p}_{KS}$ close to $100\% $ is just as 
improbable, but would indicate an exceptionnal agreement between
the theoretical distribution and the observed one.
\begin{figure} 
\psfig{file=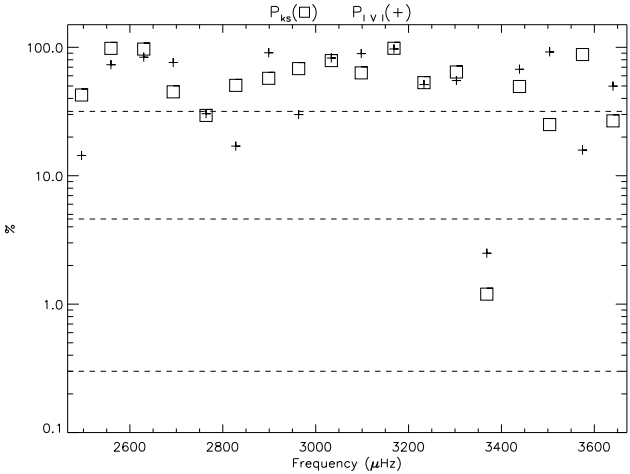,width=\columnwidth}
\caption[]{Result of the Kolmogorov-Smirnov test (square) and 
variance 
test (plus) of the modes $l=0$ and $l=1$, $17\le n \le 25$. As a 
reference,
the horizontal dashed lines delimit the upper regions which should 
contain 
$68.3\% $, $95.4\% $, $99.7\% $ of the events, respectively.}         
\label{ks9}    
\end{figure}
\par (iii) Autocorrelation of the artificial exponential 
distributions used in 
the Montecarlo method.\\
All of the p modes selected are autocorrelated over a timescale 
comparable to their damping time (2 to 4 days), usually deduced 
from the Full Width at Half Maximum (FWHM) of their lorentzian fit in 
the 
Fourier space. For the 
sake of accuracy, we have therefore used exponential distributions 
with comparable autocorrelation in order to compute the theoretical 
cumulative 
distributions $P^{p}_{V}$ and $P^{p}_{KS}$ in our Montecarlo 
simulations.
Each one is obtained by first creating a time series of a damped 
oscillator 
excited by a Gaussian noise, and then extracting the energy with the 
method 
described in Sect.~\ref{extractm}. The damping time of the oscillator 
is chosen 
such that it corresponds to a FWHM of $1 \mu Hz$ in the Fourier 
space. \\
The output of the tests, however, is only slightly modified if 
distributions made of {\it independent} points are used.\\
For both tests, Fig.~\ref{ks9} shows a very good agreement for the 
set 
of modes selected. As an exception, the energy of the mode $n=23$, 
$l=1$ is not 
exponentially distributed ($P_{|V|}=2.5\% $, $P_{KS}=1.20\% $).\\
Although the global shape of this distribution can be made compatible 
with an exponential distribution by adopting a mean value 
$10\% $ smaller than the estimated value ($P_{KS}= 30\%$ for $m=0.9 
m_{p}$), 
its variance is too large to be reconciled with the variance of an 
exponential distribution.\\
We have also analysed the distribution build with the 18 modes 
altogether (each mode is normalized by its mean energy). 
Even with this improved statistics of $18\time 210$ 
points, the variance and KS tests have not detected any significant 
deviation
from an exponential distribution ($P_{|V|}=60.7\% $, $P_{KS}=72.8\% 
$).
 
\section{Correlation of the individual modes\label{sectindep} } 
\subsection{Correlations two by two}
\begin{figure} 
\psfig{file=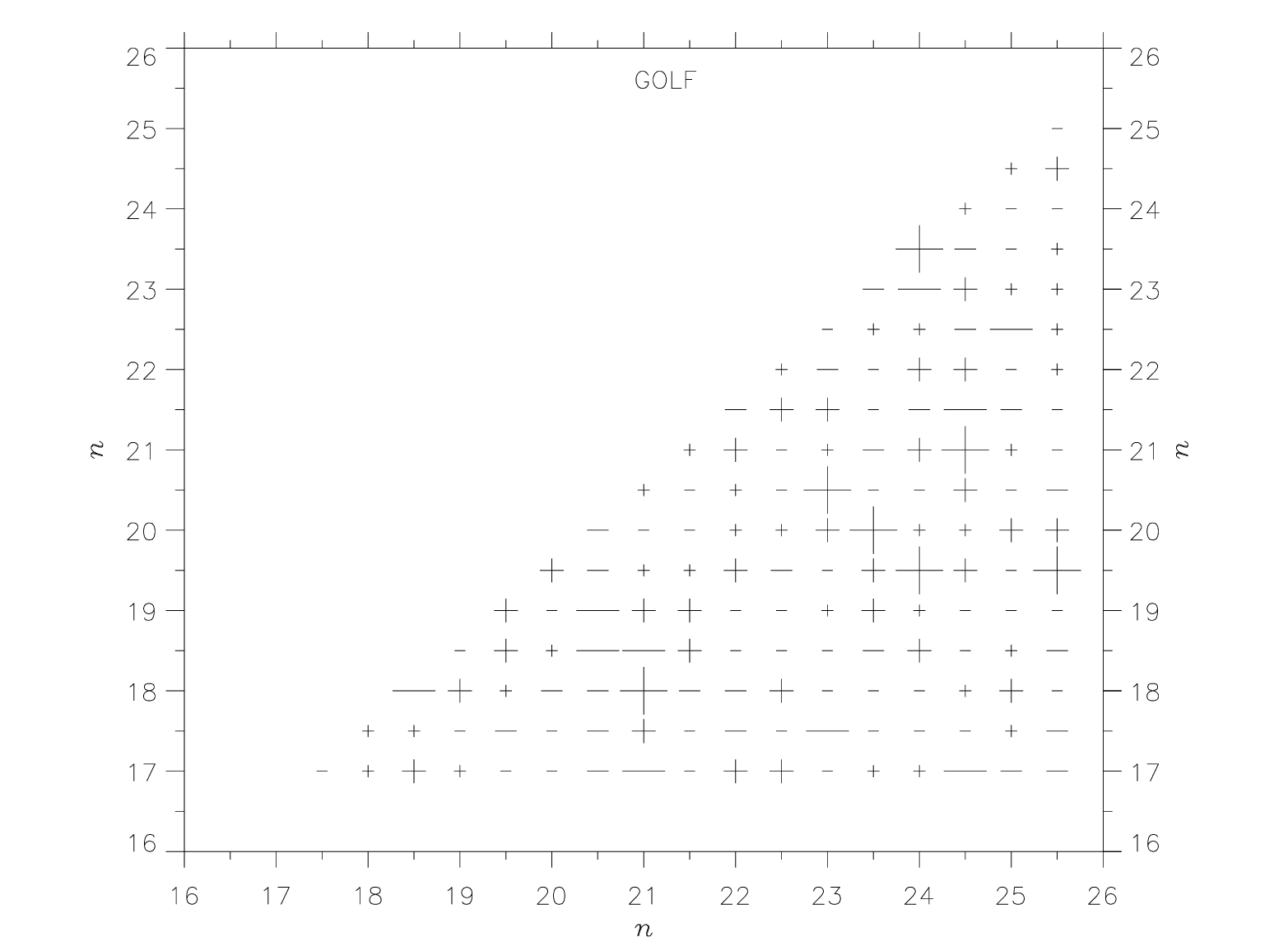,width=\columnwidth}
\caption[]{Correlation coefficients of the modes $17\le n \le 25$ 
observed by GOLF, two by two. For each value of $n$, the long and 
short ticks correspond to $l=0$ and $l=1$, respectively. The 
symbol $+$ (resp. $-$) is used for a positive (resp. negative) 
correlation.
The smallest symbols correspond to correlations smaller than the 
statistical error, intermediate and big symbols correspond to 
correlations smaller than 2 and 3 statistical errors respectively. 
}        
\label{corgolf}    
\end{figure}  
No striking general correlation appears when looking at the set 
of 18 modes displayed on Fig.~\ref{etl01}. Nor does it stand out from 
the computation of the correlations of these mo\-des, two by two. 
Although some large correlations are measured ($+17\% $ between the 
modes 
$n=20$, $l=1$ and $n=23$, $l=0$), even larger anticorrelations 
are also found ($-20\% $ between the modes $n=21$, $l=1$ and $n=24$, 
$l=1$). 
No general trend is visible, the mean value being $+0.17\%$ 
(Fig.~\ref{corgolf}).\\
The statistical error of the estimator of the correlation 
coefficient between exponential distributions, from a sample of $p$ 
points, scales like $p^{-1/2}$. For our sample of $210$ points, no 
effect smaller than $7 \% $ can therefore be detected.
Altogether, $5.5\%$ of the couples (17 out of 306 couples) present 
correlations 
contained, in absolute value, between 2 and 3 standard deviations, 
which is not 
very significant ($4.3\%$ would be expected for a normal distribution 
of the statistical error).\\
Nevertheless, a more sensitive indicator can be cons\-truc\-ted, in 
order
to determine the mean correlation coefficient more accurately.\\

\subsection{Test of the null hypothesis. Comparison with a Gamma 
distribution\label{method}}

If $k$ distributions are independent (null hypothesis), the variance 
of their sum 
should be equal to the sum of their variances. This test was 
performed 
by Baudin \etal (1996) with IPHIR data, who normalized the 
distribution by their level of noise, and found some discrepancy.\\
A fundamental feature of our method is the use of the 
exponential nature of each individual energy distribution, in order 
to compute 
the standard deviation of our estimate of the correlation, and
therefore the confidence level of our conclusions.\\
We denote by $\somme_{k,p}$ the sum of $k$ distributions of energy, 
made of $p$ events, where each of the distributions is normalized by 
its
estimated mean energy.
Since each distribution appears to be exponential within the 
statistical 
error  (Section 2), $\somme_{k,p}$ ought to resemble a 
Gamma-distribution of 
order $k$ (denoted by $\Gamma_{k}$) if they are independent, or an 
exponential
distribution of mean value $k$ if they are all identical. 
The null hypothesis can therefore be tested by comparing the observed 
distribution $\somme_{k,p}$ with the theoretical 
$\Gamma_{k}$-distribution, using the variance and KS tests. \\
Denoting by ${\cal C}_{i,j}$ the correlation coefficient between the 
modes $i$ and $j$, and ${\cal C}$ their mean value, 
the variance of $\somme_{k}\equiv\lim_{p\to\infty}\somme_{k,p}$ is 
directly 
related to these correlations:
\begin{eqnarray}
{\rm var}(\somme_{k})&=&k+2\sum_{1\le i<j\le k}{\cal 
C}_{i,j}.\label{varcor}\\
&=&k+k(k-1){\cal C}.\label{varcor2}
\end{eqnarray}
If the modes are independent, the standard deviation $\sist^{0}$ of 
the 
variance estimator of $\somme_{k}=\Gamma_{k}$ is
\begin{equation}
\sist^{0}\sim \left(2+{6\over k}\right)^{1\over2}{k\over p^{1\over2}}.
\label{sist0}
\end{equation}
Consequently, another way of testing the null hypothesis is the 
comparison of
the variance of $\somme_{k,p}$ with var$(\Gamma_{k})=k$, 
in units of the statistical error $\sist^{0}$.\\
Even if the $k$ distributions defining $\somme_{k,p}$ are 
independent and exponential, an additional error of the order of 
$p^{-1/2}$ is 
introduced in the estimation of the mean value of each exponential 
distribution.
We use a Montecarlo method of $10^{5}$ trials made from independent 
exponential distributions, in order to define the cumulative 
distributions
$P^{k,p}_{V}$ for the outcome of the variance test, and
$P^{k,p}_{KS}$ for the outcome $d(\Gamma_{k},\somme_{k,p})$ of the KS 
test.
Here again, we have used autocorrelated exponential distributions in 
the Montecarlo simulations.\\
Eq.~(\ref{varcor2}) indicates that the variance of the distribution 
gives a direct measure of the mean correlation among the modes:
\begin{equation}
{\cal C}={{\rm var}(\somme_{k,p})-k\over k(k-1)}
+{\cal O}\left(\sist\over k(k-1)\right).\label{delvar}
\end{equation}
This formulae, however, is not directly useful without an expression 
of the statistical error $\sist$ associated with the estimator of the 
variance.
If the modes are correlated, computing it requires some 
additionnal assumptions about the properties of the correlation 
(Sect.~\ref{lambhyp}). Nevertheless, $\sist$ coincides with 
Eq.~(\ref{sist0}) 
to first order. 
Together with Eq.~(\ref{delvar}), the smallest 
correlation detectable with this method scales as follows:
\begin{equation}
{\cal C}_{\rm min}\equiv {\sist^{0}\over k(k-1)}
\sim {1\over k}\left({2\over p}\right)^{1\over2},
\label{lamin}
\end{equation}
which is a factor $\surd 2/k$ smaller than the sensitivity of the 
correlation 
coefficient two by two. A better sensitivity is therefore obtained by 
summing a large number of modes. However, the hypothesis of
a constant correlation between the modes might be
questionable if the range of frequencies is large, especially since 
the mean energy and the lifetime of the modes vary significantly with 
frequency.
\subsection{GOLF results}
Both tests are of course very sensitive to the presence of a gap in 
the 
data. If $x\;\%$ of the data were filled with zeros due to an 
interruption of the instrument, the variance of $\somme_{k,p}$ would 
be 
increased by a factor $\sim (1+k)x\;\%$. Consequently, we have 
carefully 
removed from our samples the points corresponding to these gaps.\\
\begin{figure} 
\psfig{file=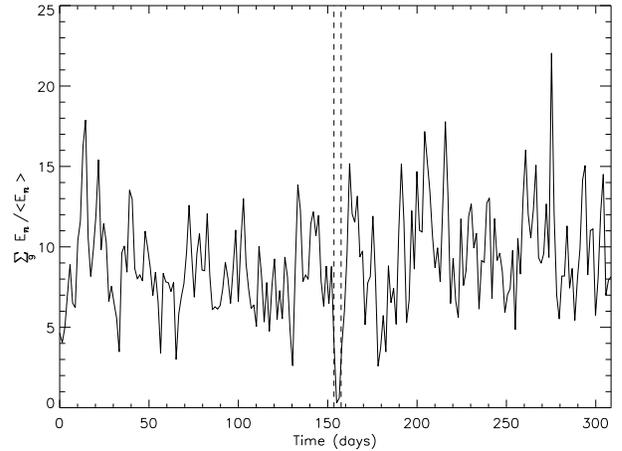,width=\columnwidth}
\caption[]{Time evolution of $\somme_{9,210}$, the sum of the 
normalized energies of 
the $9$  modes $l=0$, $17\le n \le 25$, observed by GOLF.}        
\label{gaml0t9}    
\end{figure}
\begin{figure} 
\psfig{file=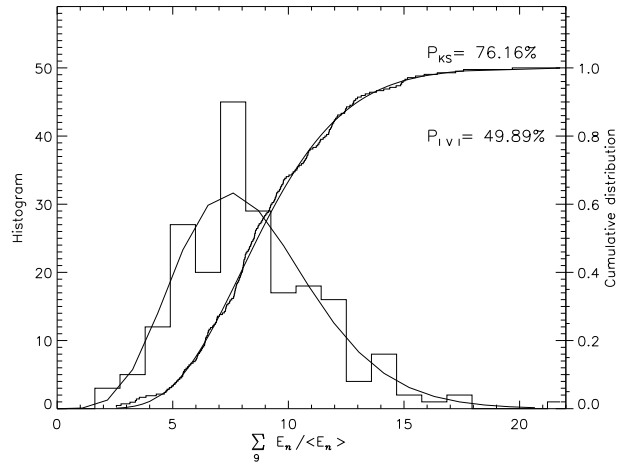,width=\columnwidth}
\caption[]{Histogram (20 bins) of the energy and cumulative 
distribution of 
the sum of the $9$  modes $l=0$, $17\le n \le 25$ observed by GOLF, 
compared to a $\Gamma_{9}$  distribution.}        
\label{gaml0d9}    
\end{figure}
The sum of the energies of the 9 modes $l=0$, $17\le n \le 25$, 
normalized 
to their mean energy, is shown in Fig.~\ref{gaml0t9}. We note in 
passing that the clear gap in the data appearing around the 156th day 
confirms 
the validity of our procedure for the extraction of the energy. As 
before, 
four days of signal have been removed from our statistical study to 
account 
for the stabilisation of the instrument, resulting in a sample made 
up of 210 points. Their distribution is successfully compared to a 
$\Gamma_{9}$-distribution in Fig.~\ref{gaml0d9} ($P_{|V|}=49.9\% $ 
and $P_{KS}=76.2\% $). The same test 
performed on the 9 modes $l=1$, $17\le n \le 25$, obtains
$P_{|V|}=31.6\% $ and $P_{KS}=34.2\% $. Applied to these 18 
modes altogether, the tests confirm again the null hypothesis 
($P_{|V|}=95.8\% $ and $P_{KS}=66.5\% $).\\
The correlation being low, we may use Eq.~(\ref{delvar}) with the 
statistical error given by Eq.~(\ref{sist0}), and obtain a 
confirmation of 
the absence of correlation, with an error bar: ${\cal C}=0.7\pm 1.4 
\% $ for 
9 modes $l=0$, and ${\cal C}=-0.1\pm 0.6\% $ for 18 modes $l=0,1$.

\subsection{Test of the ``$\lambda$-hypothesis''. Correlation due to 
an 
additive common signal\label{lambhyp}}
A refined estimate of the correlation can be obtained by making some 
assumptions about its origin. 
We build in Appendix B a simple model where the excitation is a 
mixture of 
two types of sources, which, taken separately, would result in 
uncorrelated/highly correlated modes energies respectively.\\
The first type represents the granules, which produce so many 
excitations per damping time that the correlation among the modes 
energies is close to zero.\\
The second type is hypothetical. It could be produced by some 
isolated events, possibly of magnetic origin, separated by a time 
comparable to or larger than the damping time of the modes 
considered.\\
We assume that the mode response to an excitation is linear, and 
therefore the  response to a mixture of sources is a superposition of 
the answers to the two types separately. Our model depends on a 
single parameter $\lambda$, namely the fraction of the energy of each 
mode due to the second type of sources.\\
We define in Appendix B the theoretical distribution function 
$\Gamma_{k}^{\lambda}$ corresponding to such correlated modes 
energies.
If the model is applicable, the distribution $\somme_{k,p}$ ought to 
converge, for $p\to\infty$, towards a well defined distribution 
denoted by 
$\Gamma_{k}^{\lambda}$, such that:
\begin{eqnarray}
\Gamma_{k}^{\lambda=0}(x)&=&\Gamma_{k}(x),\\
\Gamma_{k}^{\lambda=1}(x)&=&{1\over k}\e^{-{x\over k}}.
\end{eqnarray}
The variance and KS test can therefore be used, for various values 
of $\lambda$, in order to test this ``$\lambda$-hypothesis''.\\
With the definition of Appendix B, the correlation coefficient 
${\cal C}$ is related to  the coefficient $\lambda$ as follows: 
\begin{equation}
{\cal C}=\lambda^{2}.
\end{equation}
\begin{figure} 
\psfig{file=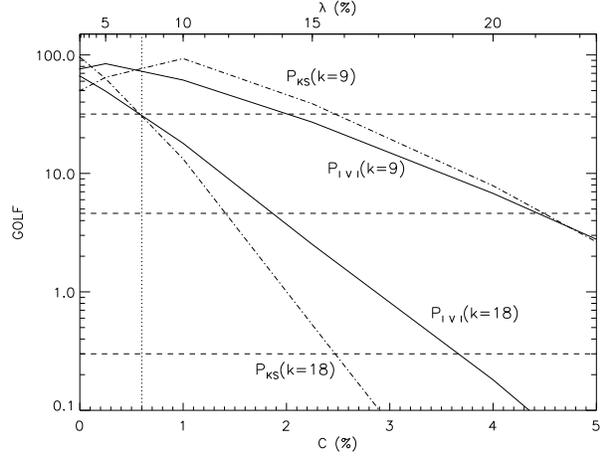,width=\columnwidth}
\caption[]{Comparison of GOLF data ($p=210$ points) with a 
theoretical 
distribution made of $k$ modes with a uniform correlation ${\cal C}$,
using the variance and KS tests. The vertical dotted line
delimits the sensitivity limit (${\cal C}_{min}=0.6\%$) defined by 
Eq.~(\ref{lamin}) for $k=18$ modes. It coincides with the value below 
which the 
corresponding variance and KS tests remain inside the upper $68.3 \%$ 
region.}        
\label{figogolf}    
\end{figure}
The statistical error $\sist$ associated with the estimator of the 
variance 
also depends on $\lambda$ according to Eq.~(\ref{astat}). 
Eq.~(\ref{delvar}) can then be used to determine the mean correlation 
${\cal C}$, with a consistent statistical error. \\
We do not expect higher order moments of the distribution $S_{k}$ to 
be more 
sensitive to a correlation between the modes, since we prove in 
Appendix B that they also vary like $\lambda^{2}$ at first order.\\
We also demonstrate that the shape of the cumulative distribution 
varies like 
$\lambda^{2}$. The sensitivity limit of the KS test is therefore 
expected to be comparable to the sensitivity limit of the variance 
test.\\
Here again, normalizing the distributions by their estimated mean 
value introduces a bias, which we take into account using a 
Montecarlo method. 
For each value of $k,\lambda$ considered, we compute from $10^{6}$ 
trials the theoretical distribution $\Gamma_{k}^\lambda$, and use 
$10^{5}$ 
other trials to define the cumulative distributions 
$P^{\lambda,k,p}_{V}$ 
and $P^{\lambda,k,p}_{KS}$ which are used for our variance and KS 
tests. For the sake of simplicity, the effect of the autocorrelation 
of each
mode is neglected here. Indeed, we know from Sect.~\ref{statest} and 
\ref{method}
that it introduces very small corrections on the cumulative 
distributions 
$P^{k,p}_{V}$ and $P^{k,p}_{KS}$.\\
We define the error bar of the correlation coefficient as the range 
of values of 
${\cal C}=\lambda^{2}$ within which the test ($P_{|V|}$ or $P_{KS}$) 
remains 
inside the upper $68.3 \%$ region, by analogy with the statistics of 
normal 
distributions.\\
Fig.~\ref{figogolf} shows that the variance and KS tests, applied to 
18 modes for various values of $\lambda$, stays within the upper 
$68.3 \%$ region for ${\cal C}\le 0.6 \% $, which coincides with the 
statistical limitation expressed by Eq.~(\ref{lamin}).\\
Nevertheless, while the measurements made by GOLF are 
compatible with a total lack of correlation of the modes, our tests 
cannot exclude that up to $\lambda_{\rm max}\sim 8 \% $ of the energy 
is common 
to the modes.

\section{Comparison with IPHIR data\label{sectiphir}}
\subsection{Extraction of 11 modes} 
\begin{figure} 
\psfig{file=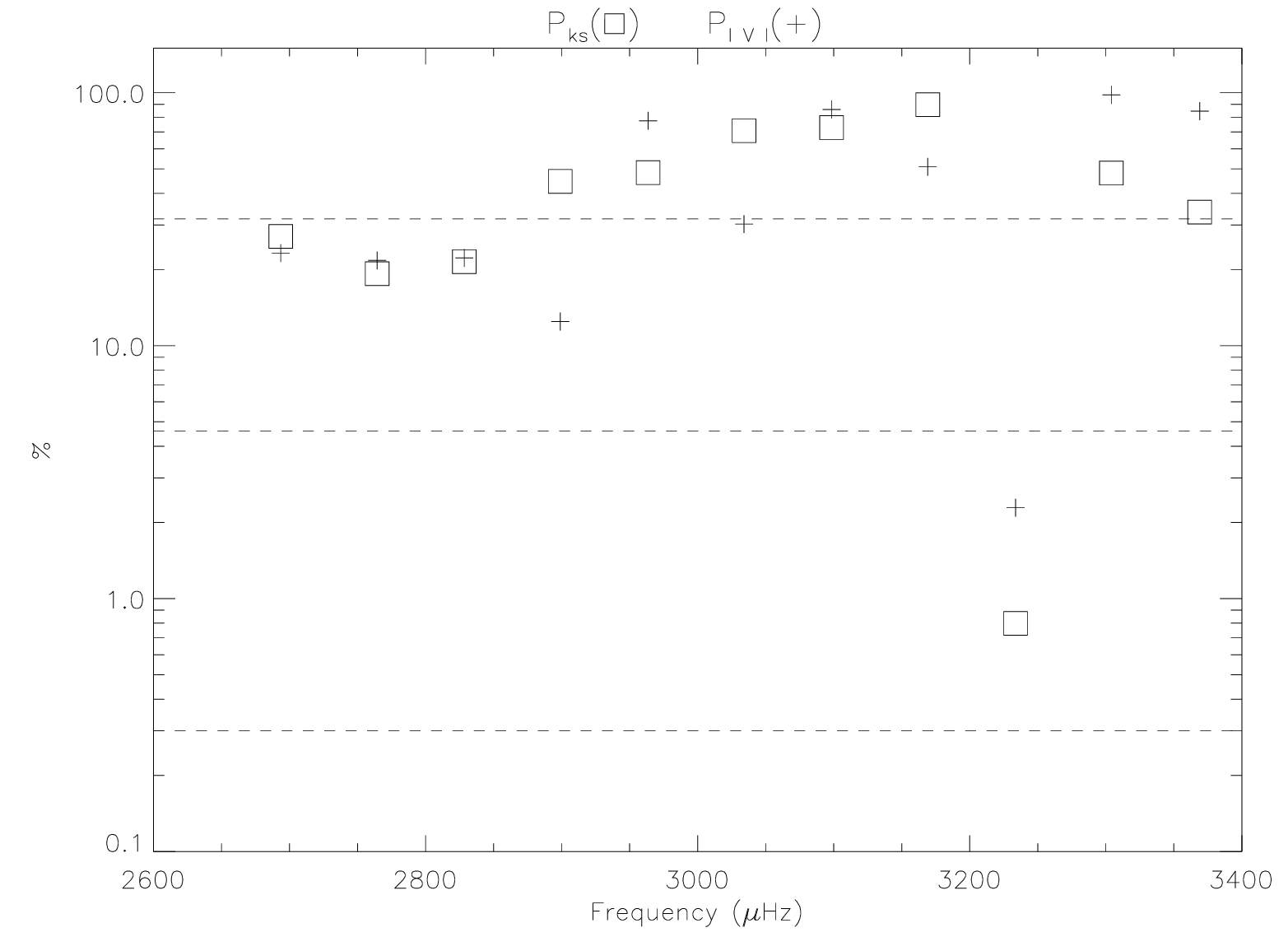,width=\columnwidth}
\caption[]{Variance test (plus) and KS test (square) for the 11 modes 
$l=0$, $19\le n \le 23$, 
and $l=1$, $18\le n \le 23$ observed by IPHIR with a 
filtering window of 6 $\mu$Hz.}        
\label{gamiph}    
\end{figure}
Since the conclusions of Baudin \etal (1996) about 160 days of IPHIR 
data were 
obtained from time variations of the power instead of the energy, 
using a 
different normalization and a different time resolution, we have 
first 
re-analysed these
data with the method described above, using the same 11 modes ($l=0$, 
$19\le
n \le 23$, and $l=1$, $18\le n \le 23$). The central frequency 
$\omega_{0}$
is  taken from Toutain \& Fr\" ohlich (1992). The higher level of 
noise limits
the size  of the filtering window to 
$6\mu$Hz, leading to a time resolution of $1.9$ days, and a 
statistical study on $82$ points. We have removed the data 
surrounding 
two gaps in the series, around the 5th and the 61st day. The 
resulting sample is shortened to $78$ points only.\\ 
According to Fig.~\ref{gamiph}, the distribution of energy of each of 
the 11 modes of IPHIR is compatible with an exponential distribution 
(apart from the mode $l=1$, $n=22$).\\
\begin{figure} 
\psfig{file=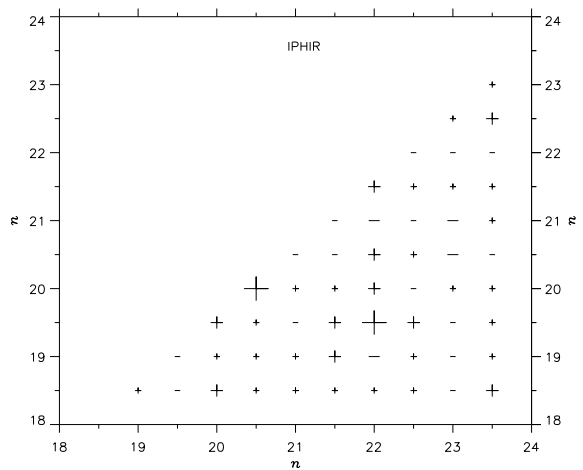,width=\columnwidth}
\caption[]{Correlation coefficients of the 11 modes observed by 
IPHIR, two by two.
For each value of $n$, the long and short ticks correspond to $l=0$ 
and $l=1$, 
respectively. The symbol $+$ (resp. $-$) is used for a positive 
(resp. negative) 
correlation. The smallest symbols correspond to correlations smaller 
than the 
statistical error, intermediate and big symbols correspond to 
correlations smaller than 2 and 3 statistical errors 
respectively.}        
\label{coriphir}    
\end{figure}
The correlation coefficient of the modes energy, two by two, is shown 
in Fig.~\ref{coriphir}. The mean value is $3.9\%$, the statistical
error being $11.3\%$. Altogether, $4.4\%$ of the couples (2 out of 45 
couples)
present correlations contained, in absolute value, between 2 and 3 
standard 
deviations, which is comparable to the $4.3\%$ expected for a normal 
distribution of the statistical error.
\subsection{Test of the null hypothesis}
We have also extracted the same modes, with the same filtering 
window, from the 
first  153 days of GOLF data to obtain a comparable sample of 78 
points.
\begin{figure} 
\psfig{file=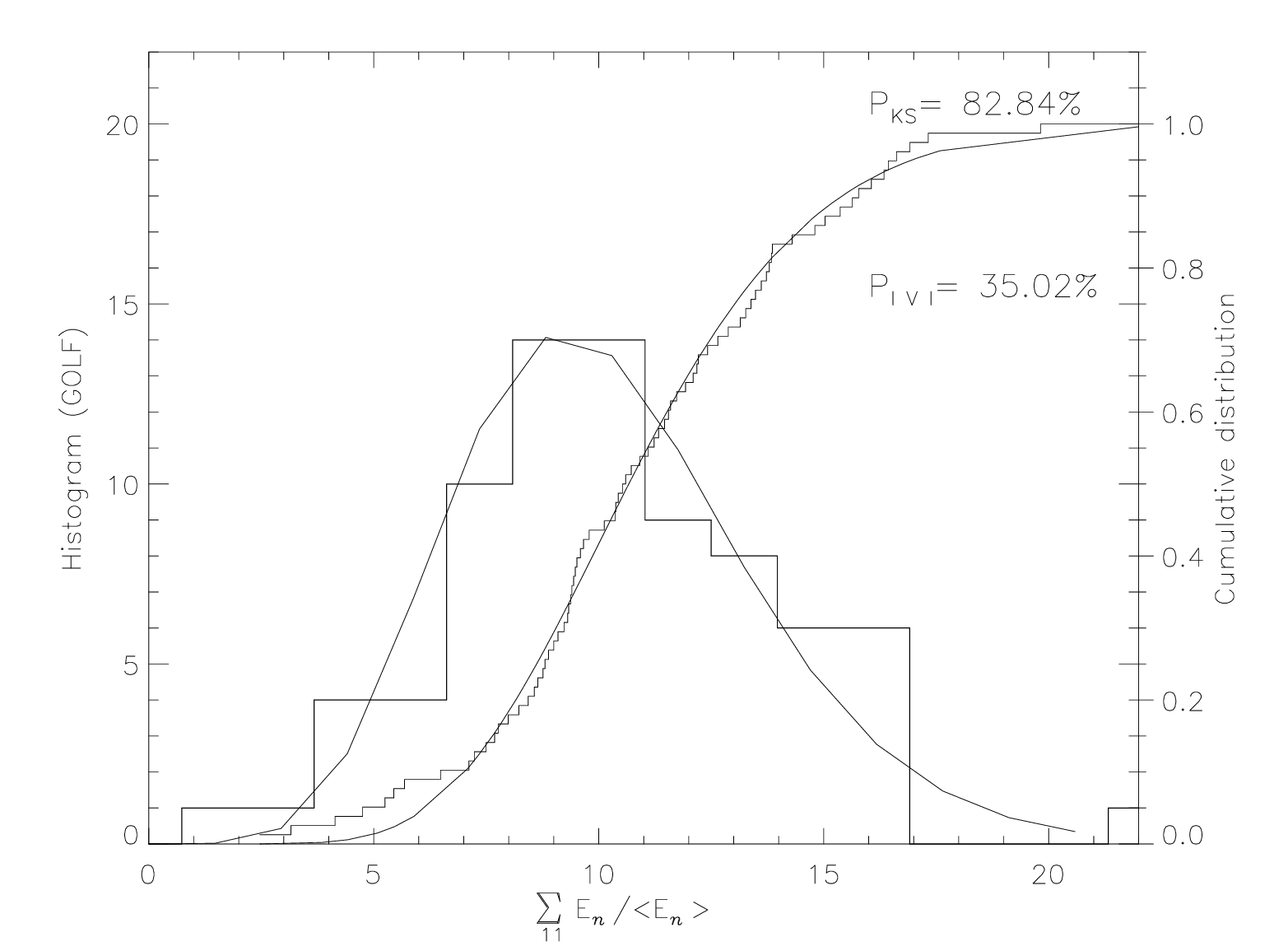,width=\columnwidth}
\psfig{file=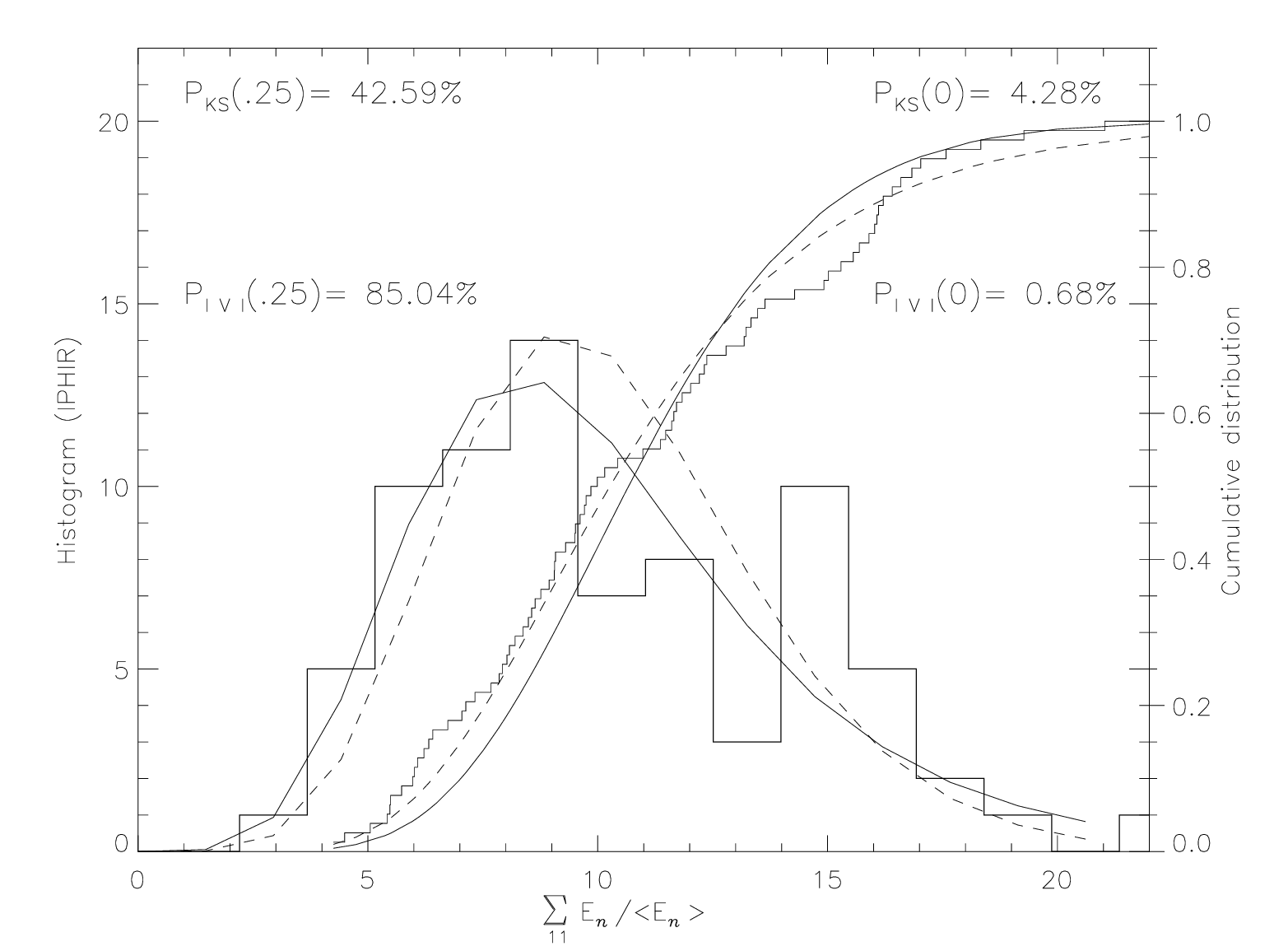,width=\columnwidth}
\caption[]{Histogram (15 bins) of $\somme_{11,78}$ and cumulative 
distribution, 
compared  to a $\Gamma_{11}$ distribution for GOLF (above) and IPHIR
(below). In the IPHIR case, the dashed lines correpond to the 
theoretical 
distribution $\Gamma_{11}^{\lambda=.25}$.}        
\label{gamgo}    
\end{figure}
The difference between the two series appears on the distribution
$\somme_{11,78}$ shown in Fig.~\ref{gamgo}, where both the variance 
and the 
KS tests indicate that the modes are likely to be 
less independent in IPHIR data than in GOLF data.\\
The tests applied to GOLF data are compatible with the null 
hypothesis ($P_{|V|}=35.0\%$, $P_{KS}=82.8\%$), which is consistent 
with the results of Sect.~\ref{sectindep}.\\
By contrast, the same tests applied to IPHIR data {\it reject the 
null 
hypothesis} with a $99.32\%$ confidence level with the variance test, 
and a $95.72\%$ confidence level for the KS test
($P_{|V|}=0.68\%$, $P_{KS}=4.28\%$).

\subsection{Test of the ``$\lambda$-hypothesis''}
\begin{figure} 
\psfig{file=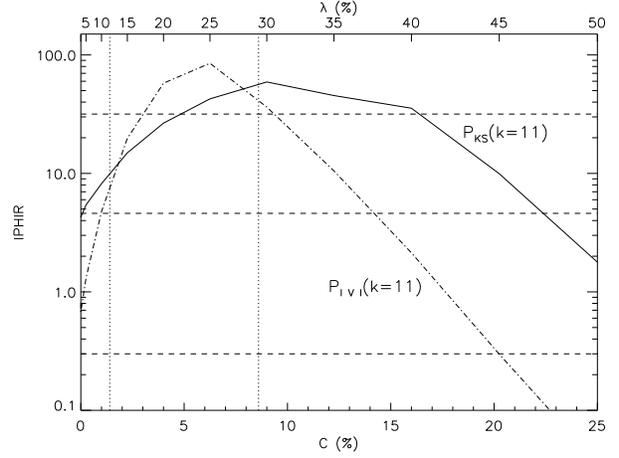,width=\columnwidth}
\caption[]{Comparison of IPHIR data ($p=78$ points) with a 
theoretical 
distribution made 
of $k=11$ modes containing a fraction $\lambda={\cal C}^{1/2}$ of 
energy in 
common. The vertical dotted lines delimit 
the range of the correlation coefficient determined by 
Eq.~(\ref{delvar}). 
The KS test remain inside the upper $68.3 \%$ region
for ${\cal C}=10.7\pm 5.9\% $, corresponding to a fraction 
$\lambda=31.3\pm 9.4\% $.}        
\label{figophir}    
\end{figure}

While the cumulative distribution $\somme_{11,78}$ of GOLF
does not show any systematical trend when compared to $\Gamma_{11}$,
the cumulative distribution $\somme_{11,78}$ of IPHIR shows a clear 
trend. 
This trend is successfully suppressed  when compared to the 
distribution
$\Gamma_{11}^{0.25}$ (Fig.~ \ref{gamgo}). Fig.~\ref{figophir} shows 
that
within our simple model, the signal of IPHIR would be absolutely 
normal as regards our tests ($P_{|V|}=85.0\%$, $P_{KS}=42.6\%$) 
if a fraction $\lambda=25\%$ of each mode energy were common to all 
the modes, 
correponding to a mean correlation ${\cal C}=6\%$.\\
Error bars are obtained by varying the parameter $\lambda$: the 
variance test 
leads to ${\cal C}=6.1\pm 3.3\%$, while the KS test obtains 
${\cal C}=10.7\pm 5.9\%$.\\
Moreover, the correlation computed from Eq.~(\ref{delvar}) with the 
statistical error given by Eq.~(\ref{astat}) is ${\cal C}=5.0\pm 3.6 
\% $.
Although this analytical estimate is less reliable than the tests 
based on Montecarlo simulations (Eq.~(\ref{astat}) neglects the error 
in estimating the mean energy of each mode), it is useful as a quick 
check of the results.\\
It is therefore comforting to notice, as can be seen in 
Fig.~\ref{figophir},
that the range of correlations defined by these three methods overlap 
in the range $4.8\%<{\cal C}<8.6\%$, correponding to 
$21.9\%<\lambda<29.3\%$.
Of course, this overlapping region cannot be directly interpreted in 
terms of a standard deviation. We shall adopt the conservative range 
obtained with the KS test, which takes the full distribution into 
account: 
${\cal C}=10.7\pm 5.9\% $, corresponding to a fraction 
$\lambda=31.3\pm 9.4\% $.
\subsection{Additionnal checks}
In order to check the possibility that the correlation might come 
from a multiplicative noise (such as due to a pointing noise), we 
have 
computed the correlation between $11$ windows of noise centered 
$20\mu 
Hz$ (resp. $27\mu Hz$) to the right of each mode.  
This test indicates that the noise itself is not correlated, with
$P_{|V|}=77.2\% $ and $P_{KS}=25.5\% $ 
(resp. $P_{|V|}=68.4\% $ and $P_{KS}=34.3\% $).\\
We have checked the effect of changing the size of the filtering 
window 
to 4 $\mu$Hz 
(no noise, but low statistics of 52 points) and 8 $\mu$Hz (good 
statistics 
of 106 points, but IPHIR is influenced by the noise). While a smaller 
filtering 
window still favours $\lambda\sim 25\%$, a larger window takes into 
account a 
significant fraction of uncorrelated noise, as 
expected, resulting in a slightly lower value of $\lambda$.\\
One might also suspect that the discrepency between the IPHIR 
distribution 
$\somme_{11,78}$ and a $\Gamma_{11}$ distribution is due to the mode 
$l=1$, $n=22$ which is not well fitted by an exponential 
distribution (see Fig.~\ref{gamiph}). Nevertheless, performing the 
same 
analysis 
without this particular mode leads to the same conclusion:
$P_{|V|}=0.28\% $ and $P_{KS}=6.9\% $ if $\lambda=0$, while 
$P_{|V|}=78.2\% $ and $P_{KS}=38.5\% $ if $\lambda=25\%$.

\section{Conclusion} 
The exponential nature of the energy distribution of each mode has 
been 
used to compute their mean correlation coefficient with a consistent 
error bar.\\
Two tests based on Montecarlo simulations, and one analytical 
formulae, 
applied to 310 days of GOLF data, support the null hypothesis of no 
mean correlation among the modes $17\le n\le 25$, $l=0,1$, 
with an accuracy of ${\cal C}=0\pm 0.6\% $.\\
Our analysis of the modes correlation in IPHIR data,
using these statistical tools, gives an accurate 
statistical support to the tentative conclusions of Baudin \etal 
(1996).
The variance test rejects the null 
hypothesis with a $99.3\% $ confidence level.\\
The presence of a clear correlation among p-mode energies in IPHIR 
data strongly constrains the standard picture of stochastic 
excitation. If really of solar origin, it suggests the existence of 
an additional source of excitation, other than the granules. We have 
built a one parameter model of random excitations separated by a time 
comparable to the damping time of the modes, added to the usual 
granule excitations. IPHIR data are fully compatible with this 
``$\lambda$-hypothesis'', if a fraction $\lambda=31.3\pm 9.4\% $ of 
each mode energy is due to this additionnal source of excitation, 
resulting in a mean correlation ${\cal C}=10.7\pm 5.9\% $ among the 
modes.\\
On the other hand, the absence of correlation in GOLF data support 
the standard picture of stochastic excitation by the granules only.\\
This difference between IPHIR and GOLF data can be interpreted as a 
change
from $\lambda=31.3\pm 9.4\% $ in IPHIR data to less than $8\%$ in 
GOLF data.\\
This evolution could be related to the change in magnetic activity, 
since 
the GOLF data correspond to a period close to the solar minimum while 
the IPHIR data correspond to a period closer to the solar maximum.
If this is true, a confirmation will be obtained by 
performing this same analysis on GOLF data when we approach the 
solar maximum, in a couple of years. VIRGO data will also 
be useful in order to identify the possible role of the measurement 
techniques 
(velocity/intensity) in the determination of the correlation. \\
However, the mechanism by which the magnetic field influences the 
excitation 
of the modes, \ie the nature of these hypothetical exciting events 
remains to be explored in more detail.

\acknowledgements
We thank Claus Fr\" ohlich and Thierry Tou\-tain for providing access 
to IPHIR 
data cleaned of satellite pointing noise. We are grateful to Michel 
Tagger and Romain Teyssier for many useful discussions, and Thierry 
Appourchaux for constructive comments about the manuscript. SOHO is a 
mission
of international cooperation between ESA and NASA.
\appendix

\section{Extraction of the time evolution of the energy 
\label{apphilbert}}

Let us consider the real displacement $y(t)$ filtered through 
a double window with a width $\Delta \omega$, centred on the 
frequencies 
$\pm \omega_{0}$:
\begin{equation}
y(t)\equiv 
\int_{\omega_{0}-{\Delta \omega\over 2}}^{\omega_{0}+{\Delta 
\omega\over 2}}
{\hat y}(\omega)\e^{i\omega t}\d \omega +\cc ,\label{yt}
\end{equation}
where $\cc$ is the complex conjugate of the first term. 
By analogy with an oscillator of eigenfrequency $\omega_{0}$, the 
energy is 
defined as the sum of a kinetic and a potential part:
\begin{equation}
{E\over M}(t)\equiv {1\over2}(v^{2}+\omega_{0}^{2}y^{2}).
\end{equation}
From Eq.~(\ref{defg}) and (\ref{yt}), the filtered displacement and 
velocity can be written as follows:
\begin{eqnarray}
y(t)&=&{\cal R}{\rm eal}\left\{2\;\e^{i\omega_{0} t}f_{y}(t)\right\} 
,\\
v(t)&=&{\cal R}{\rm eal}\left\{2\;\e^{i\omega_{0} t}f_{v}(t)\right\} 
,\\ 
{E\over M}(t)&=&|f_{v}|^{2}+\omega_{0}^{2}|f_{y}|^{2}\nonumber \\
& &+{\cal R}{\rm eal}
\left\{\e^{2i\omega_{0} 
t}(f_{v}^{2}(t)+\omega_{0}^{2}f_{y}^{2}(t))\right\}.
\label{em}
\end{eqnarray}
We first deduce from the relation $v\equiv \d y/\d t$ that $f_{v}$ 
and 
$f_{y}$  are related as follows:
\begin{eqnarray}
f_{v}(t)&=&
i\omega_{0}\int_{-{\Delta \omega\over 2}}^{+{\Delta \omega\over 2}}
{\hat y}(\omega_{0}+\omega)\e^{i\omega t}
\left(1+{\omega\over\omega_{0}}\right)\d \omega,\\
&=&i\omega_{0}f_{y}(t)
\left\{1+{\cal O}\left({\Delta\omega\over\omega_{0}}\right)\right\}.
\end{eqnarray}
From Eq.~(\ref{defg}), the non-zero Fourier components of $f_{y}$ 
(and 
$f_{v}$) correspond to frequencies between $-\Delta \omega/2 $ and 
$+\Delta \omega/2 $.
If, as is usually the case, $\Delta \omega\ll \omega_{0}$, the high 
frequency  oscillations of $\exp \pm 2i\omega_{0}t$ are well 
separated from
the slower  variations of $f_{y}(t)$ and $f_{v}(t)$, and 
Eq.~(\ref{em}) is
approximated by Eq.~(\ref{energ}).

\section{One parameter model of correlated modes \label{asens}}
The amount of energy which is coherent among the 
modes can be estimated by constructing a simple one-parameter model 
as follows. 
Indexing the modes by $j\in\{1,k\}$, we assume that the velocity 
residual of
each mode $V_{j}(t)=v_{j}(t)+\alpha_{j}v_{0}(t)$ is made of a 
superposition 
of two independent signals, where $v_{0}(t)$ is common to all the 
modes, and 
all the $v_{j}(t)$, $j\in\{0,k\}$, are independent.\\
Using the filtering method described in Section 1, we introduce 
the parameter $\theta_{j}$ as: 
\begin{eqnarray}
f_{V_{j}}(t)&=& f_{v_{j}}(t)+\alpha_{j}f_{v_{0}}(t)\\
&\equiv&{\sigma(f_{V_{j}})\over 2}\left(
\cos\theta_{j}\left|\begin{array}{l}  
r_{j}\\i_{j}
\end{array}\right.
+\sin\theta_{j}\left|\begin{array}{l}  
 r_{0}\\i_{0}
\end{array}\right.\right),
\end{eqnarray}
where $(r_{j},i_{j})$, $j\in\{0,k\}$, are independent normalized 
normal 
distributions. The quantity $\lambda_{j}\equiv\sin^{2}\theta_{j}$ can 
be 
interpreted as the ratio of 
the energy in the common signal to the total energy of the mode.
We denote by $e_{j}$ the energy of the signal filtered in the
Fourier  space, normalized to its mean value:
\begin{equation}
e_{j}\equiv {1\over2}\left\{(r_{j}\cos\theta_{j} + 
r_{0}\sin\theta_{j})^{2}+( 
i_{j}\cos \theta_{j}+ i_{0}\sin\theta_{j})^{2}\right\}.
\label{ej}
\end{equation}
The correlation between two modes ${\cal C}_{i,j}$ is then:
\begin{equation}
{\cal C}_{i,j} = \lambda_{i}\lambda_{j}.
\end{equation}
For the sake of simplicity, $\lambda_{j}\equiv\lambda $ is assumed to 
be
independent of the mode $j$. This is equivalent to assuming that the 
correlation is uniform among the modes.\\ 
The sum $\somme_{k}$ of the normalized energies is defined as:
\begin{equation}
\somme_{k}\equiv \sum_{j=1}^k e_{j}.\label{defS}
\end{equation}
The variance of the estimator of the variance depends on the fourth 
moment of the distribution, and is equal to:
\begin{eqnarray}
\sist^{2}={2k(k+3)\over p}+\nonumber \\
{2k(k-1)\lambda^{2}\over 
p}[2(k+9)+12(k-2)\lambda+(4k^2-10k+9)\lambda^{2}].
\label{astat}
\end{eqnarray}
Let us show that the higher order moments $\mu_{l}$ of the 
distribution 
$\somme_{k}$ vary like $\lambda^{2}$, to first order.\\
\begin{equation}
\mu_{l}\equiv \left\langle(\somme_{k}-k)^l\right\rangle,\label{mup}
\end{equation}
where $\langle\;\rangle$ denotes the expectation value of the 
distribution.
As the transformation $\theta\to-\theta$ does not change the 
distribution defined by Eq.~(\ref{ej}), only the even powers of 
$\sin\theta$
can contribute to $\mu_{l}$. It can therefore be expanded into powers 
of $\lambda=\sin^{2}\theta$. Let us develop the product in 
Eq.~(\ref{mup}) and prove that the term of order $\lambda$ is 
zero. We use below the special relation between the centred moments 
$\tilde\mu_{l},\tilde\mu_{l-1},\tilde\mu_{l-2}$ of 
a $\Gamma_{k}$-distribution of order $k$:
\begin{eqnarray}
\mu_{l}&=& \tilde\mu_{l}+A\lambda+{\cal O}(\lambda^{2}),\\
A&=&\left\langle
{l(l-1)\over2^{l-1}}\left(\sum_{j=1}^k 
r_{0}r_{k}+i_{0}i_{k}\right)^{2}
\left(\sum_{j=1}^k 
r_{k}^{2}+i_{k}^{2}-2\right)^{l-2}\right\rangle\nonumber\\
& &-l\tilde\mu_{l}\\
&=&
l(l-1)\left(\tilde\mu_{l-1}+k\tilde\mu_{l-2}\right)-l\tilde\mu_{l}\\
&=&0.\label{egalzero}
\end{eqnarray}
Let us now show that the probability density $\Gamma_{k}^{\lambda}(x)$
of $\somme_{k}$ also varies like $\lambda^{2}$, to first order.\\
Given the Eqs. (\ref{ej})-(\ref{defS}) defining $\somme_{k}$, we can 
expand $\Gamma_{k}^{\lambda}(x)$ in powers of $\lambda$. 
\begin{equation}
\Gamma_{k}^{\lambda}(x)=\Gamma_{k}(x)+\lambda g(x)+{\cal 
O}_{x}(\lambda^{2}).
\end{equation}
The moment of order $l$ is defined as:
\begin{eqnarray}
\mu_{l}&\equiv&\int_{-\infty}^{+\infty}(x-k)^l 
\Gamma_{k}^{\lambda}(x)\d x\\
&=&\tilde\mu_{l}
+\lambda\int_{-\infty}^{+\infty}(x-k)^l g(x)\d x
+{\cal O}_{x}(\lambda^{2}).
\end{eqnarray}
The fact that every moment $\mu_{l}$ varies at least like 
$\lambda^{2}$
(Eq.~\ref{egalzero}) implies:
\begin{equation}
\int_{-\infty}^{+\infty}(x-k)^l g(x)\d x=0.\label{intenul}
\end{equation}
The only function $g(x)$ satisfying Eq.~(\ref{intenul}) for any value 
of 
$l$ is $g\equiv 0$. Therefore both $\Gamma_{k}^{\lambda}(x)$ and its 
primitive (\ie the cumulative distribution of $\somme_{k}$ involved 
in the KS 
test) vary like $\lambda^{2}$ to first order.\\
We conclude that the sensitivity of the KS test is comparable to the 
sensitivity of the variance test.

\end{document}